BMC Public Health

STUDY PROTOCOL

Open Access

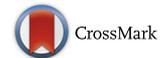

# Characterisation of exposure to non-ionising electromagnetic fields in the Spanish INMA birth cohort: study protocol

Mara Gallastegi[1,2], Mònica Guxens[3,4,5,6]*, Ana Jiménez-Zabala[1,7], Irene Calvente[5,8], Marta Fernández[9], Laura Birks[3,4,5], Benjamin Struchen[10,11], Martine Vrijheid[3,4,5], Marisa Estarlich[5,12], Mariana F. Fernández[5,8], Maties Torrent[5,13], Ferrán Ballester[5,12], Juan J Aurrekoetxea[1,7,14], Jesús Ibarluzea[1,5,7], David Guerra[9], Julián González[15], Martin Röösli[10,11] and Loreto Santa-Marina[1,5,7]

## Abstract

**Background:** Analysis of the association between exposure to electromagnetic fields of non-ionising radiation (EMF-NIR) and health in children and adolescents is hindered by the limited availability of data, mainly due to the difficulties on the exposure assessment. This study protocol describes the methodologies used for characterising exposure of children to EMF-NIR in the INMA (*Infancia y Medio Ambiente*- Environment and Childhood) Project, a prospective cohort study.

**Methods/Design:** Indirect (proximity to emission sources, questionnaires on sources use and geospatial propagation models) and direct methods (spot and fixed longer-term measurements and personal measurements) were conducted in order to assess exposure levels of study participants aged between 7 and 18 years old. The methodology used varies depending on the frequency of the EMF-NIR and the environment (homes, schools and parks). Questionnaires assessed the use of sources contributing both to Extremely Low Frequency (ELF) and Radiofrequency (RF) exposure levels. Geospatial propagation models (NISMap) are implemented and validated for environmental outdoor sources of RFs using spot measurements. Spot and fixed longer-term ELF and RF measurements were done in the environments where children spend most of the time. Moreover, personal measurements were taken in order to assess individual exposure to RF. The exposure data are used to explore their relationships with proximity and/or use of EMF-NIR sources.

**Discussion:** Characterisation of the EMF-NIR exposure by this combination of methods is intended to overcome problems encountered in other research. The assessment of exposure of INMA cohort children and adolescents living in different regions of Spain to the full frequency range of EMF-NIR extends the characterisation of environmental exposures in this cohort. Together with other data obtained in the project, on socioeconomic and family characteristics and development of the children and adolescents, this will enable to evaluate the complex interaction between health outcomes in children and adolescents and the various environmental factors that surround them.

**Keywords:** Electromagnetic fields, Radiofrequency, Extremely low frequency, Magnetic fields, Child, Adolescent, Birth cohort, Exposure assessment, Environmental exposure

* Correspondence: mguxens@creal.cat
[3]ISGlobal, Centre for Research in Environmental Epidemiology (CREAL), Barcelona Biomedical Research Park, C/Doctor Aiguader 88, 08003 Barcelona, Spain
[4]Pompeu Fabra University, C/Doctor Aiguader 88, 08003 Barcelona, Spain
Full list of author information is available at the end of the article

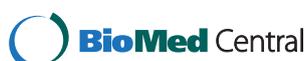

## Background

There is limited research on exposure to electromagnetic fields and its effects on children and adolescents. Both foetuses and children are especially vulnerable to persistent toxic chemicals in the environment, but evidence on health effects of electromagnetic fields of non-ionising radiation (EMF-NIR) and whether the children are more vulnerable remains unclear [1, 2].

Measuring exposure to EMF-NIR is a great challenge for researchers given, the ubiquity of the exposure, the diversity of sources, the spatial and temporal variability in emissions, and the different personal uses and behaviours in relation with the exposure sources, among other factors. There are different approaches for assessing the exposure which may be classified in indirect methods (distance to the emission sources, questionnaires and geospatial propagation models) and direct methods (spot and fixed longer-term measurements and personal measurements). The methodology also varies depending on the frequency of EMF-NIR. Usually EMF-NIR is divided in two big groups which are Extremely Low Frequency (ELF) fields and Radiofrequency (RF) fields which cover the frequency range between 0 Hz and 300 Hz and 10 MHz and 300 GHz respectively. Whilst there is a direct relation between magnetic and electric fields in RFs, there is not in ELFs due to the fact that the last are not measured in the far-field [3]. Therefore, separated measurement of magnetic (ELF-MF) and electric (ELF-EF) fields is needed when measuring ELFs. Intermediate frequencies (IFs) which cover the frequency range between 300 Hz and 10 MHz, do not tend to be analysed on their own, there being a tendency to divide the NIR spectrum into ELFs and RFs, with IFs lying at the top and bottom of these two ranges respectively. Lastly, the differences in the study protocol of studies that use direct methods is also due to the environment where the assessment is done, i.e. outdoors (parks, playgrounds, etc.) or indoors (homes, schools, etc.).

Regarding indirect methods for assessing ELF radiation, initial studies estimated the exposure to this type of magnetic field (ELF-MF) using classification systems which take into account the proximity to high voltage power lines and the size and configuration of electrical wiring among other variables [4, 5]. On the other hand, indirect methods for assessing exposure to RF fields have mainly considered the use of mobile or cordless phones [6–12], assessed using a questionnaire, or the distance between the home and the mobile phone base stations [13–15]. These indirect methodologies have been widely criticised since exposure estimated in these ways often does not correspond to levels obtained using exposimeters [16, 17], being the use of the last ones the most accurate procedure for individual exposure classification.

Spatial propagation models are another indirect method for characterising individual exposure. With regards to ELFs, Nassiri et al. [18] developed an interpolation method for estimating exposure to ELF-MFs, but the model has yet to be validated. We have not found any other geospatial model for assessing exposure to ELF-MFs and any model for assessing exposure to ELF-EFs. In contrast various propagation prediction models have been developed for estimating environmental exposure to outdoor RF sources in recent years. Some of them only included RF exposure exclusively from mobile phone base stations [19–22]. Apart from assessing exposure from mobile phone base stations, the geospatial models developed by Anglesio et al. [23] and by Bürgi et al. (NISMap) [24] also assess exposure due to radio and television transmitters. While the model proposed by Anglesio et al. (2011) tends to overestimate the exposure, the NISMap has been validated and successfully used in epidemiological studies from Switzerland [24, 25] and the Netherlands [26, 27], showing a good correlation with spot measurements conducted indoors and outdoors. However, neither the aforementioned models consider RF exposure from individual use of electronic devices (mobile or cordless phones, tablets, computers, etc.) or indoor sources, such as Wireless Access Points for WiFi technology and cordless phone base stations. This information is usually collected using questionnaires.

The characterisation of individual exposure to EMF-NIR by direct methods has been generally based on measurements in specific places at one point in time (spot), during a period of time (fixed longer-term) or by measurements using personal portable exposimeters over at least 24 h to assess the exposure of individuals in their daily life (personal measurements) [28–30]. In contrast to indirect methods, direct methods are a better approach for assess the real EMF-NIR exposure and indirect methods should be contrasted or validated by direct methods.

The INMA (INfancia y Medio Ambiente- Environment and Childhood) Project (http://www.proyectoinma.org) is an ongoing prospective population-based birth cohort study concerned with the associations between pre- and post-natal environmental exposures and child growth and development [31]. This paper describes the methodologies for the characterisation of the exposure to EMF-NIR in children from the INMA Project. This will enable us to evaluate, in a more comprehensive way, in later phases of follow-up, the complex interaction between EMF-NIR exposure and children's health and development as well as the environmental factors around them.

## Methods/Design

The characterisation of EMF-NIR exposure started in 2012. Part of the methodology described in this paper has been developed under two European projects, namely, the "Radiofrequency ElectroMagnetic fields exposure and BRAiN DevelopmenT study" (REMBRANDT Project) and "Generalised EMF research using novel methods. An integrated approach: from research to risk assessment and support to risk management" (GERoNiMO Project) (http://www.crealradiation.com).

## Study population

The study population corresponds to five out of the seven Spanish regions (Menorca, Granada, Valencia, Sabadell and Gipuzkoa) involved in the INMA Project, including a total of 1900 children and adolescents aged between 7 and 18 years old [31].

## Exposure assessment in the INMA-Cohort
### Measurement equipment

The measurements of EMF-NIR were performed with several devices, all of them properly calibrated. The specifications of each measurement device are listed in Table 1. For measuring narrow- and broadband ELF fields and broadband RF fields strength, a NBM-550 Broadband Field Meter basic unit was used in one of the study regions, with an EHP-50D Electric Field and Magnetic Field and Flux Density Isotropic Probe Analyser for fields of 5 Hz to 100 kHz and an EF 0691 Isotropic Probe for frequencies of 100 kHz to 6 GH, all from Narda Safety Test Solutions (Germany). In another study region, for broadband measurements at ELFs and RFs, a TS/001/UB Taoma Broadband Field

Meter basic unit was used with TS/002/BLF and TS/003/ELF isotropic probes for analysing the magnetic and electric fields respectively, in the 15 Hz-100 kHz frequency range and a TS/004/EHF isotropic electric field probe for the 100 kHz to 6 GHz frequency range, all from Tecnoservizi (Rome, Italy) [32, 33]. For measuring narrowband RF fields in the 87.5 MHz–6 GHz range, ExpoM-RF 3 personal portable exposimeter (Fields at work; Zurich, Switzerland) was used in all the regions of the Project. In addition, the following were available: global positioning system devices (GPS), laser distance meters (Fluke 414D and Professional GLM 30, Bosch Brand), optical fiber cables to connect probes to the computer or the basic unit, and non-conducting tripods, as well as suitable software for data mining.

## Exposure characterisation

Table 2 lists the types of methodologies carried out in each of the study regions of the INMA cohort involved in this research.

1. Indirect methods for assessing EMF-NIR field exposure
   1.1. Proximity to emission sources and questionnaires on sources use

## Methodology

Information regarding characterisation of sources that contribute to both ELF and RF fields in the environments

**Table 1** Specification of the measurement devices

| Measurement devices | EMF-NIR | Frequency range | | Measurement range[a] | Manufacturer |
| | | Broadband | Narrowband | | |
| Basic unit NBM-550 | | | | | Narda Safety Test Solutions |
| Probe EHP-50D | ELF-MF/EF | 5 Hz-100 kHz[b] | Span of 100 Hz, 200 Hz, 500 Hz, 1 kHz, 2 kHz, 10 kHz, & 100 kHz[c] | 0.3 nT-100 μT 5 mV m$^{-1}$-1 kV m$^{-1}$ | |
| Probe EF-0691 | RF | 100 kHz-6 GHz | - | 0.375 V m$^{-1}$-650 V m$^{-1}$ | |
| Taoma basic unit TS/001/UB | | | | | Tecnoservizi |
| Probe TS/002/BLF | ELF-MF | 15 Hz-100 kHz | Span of 5 kHz, 10 kHz & 100 kHz | 100 nT-10 mT | |
| Probe TS/003/ELF | ELF-EF | 15 Hz-100 kHz | Span of 5 kHz, 10 kHz & 100 kHz | 10 V m$^{-1}$-100 kV m$^{-1}$ | |
| Probe TS/004/EHF | RF | 100 kHz-6 GHz | - | 0.2 V m$^{-1}$-340 V m$^{-1}$ | |
| Personal exposure meter (ExpoM-RF 3) | RF | - | 16 bands (87.5 MHz-6GHz)[d] | 0.003/0.02[e] V m$^{-1}$-5 V m$^{-1}$ | Fields at work |

ELF extremely low frequency, RF radiofrequency, MF magnetic field, EF electric field; [a]The limit of quantification is the same as the lower limit of the measurement range; [b]It is possible to perform measurements with different spans or bandwidths; [c]The start frequency for each span corresponds to 1.2 % of the span. The frequency bands are: 5–100 Hz; 5–200 Hz; 6–500 Hz; 12 Hz-1 kHz; 25 Hz-2 kHz; 120 Hz-10 kHz; 1.2-100 kHz. When in remote mode (disconnected from the control unit), 500 Hz is the minimum span that can be measured; [d]Frequency bands: 87.5 - 108 MHz; 470–790 MHz; 791 – 821 MHz; 832 – 862 MHz; 880 – 915 MHz; 925 – 960 MHz; 1710 – 1785 MHz; 1805 – 1880 MHz; 1880 – 1900 MHz; 1920 – 1980 MHz; 2110 – 2170 MHz; 2400–2485 MHz; 2500 – 2570 MHz; 2620 – 2690 MHz; 3400 – 3600 MHz; 5150 – 5875 MHz; [e]Depending on the frequency band
Additional information regarding the equipment can be found in https://www.narda-sts.com/en/; http://www.westek.com.au/wp-content/uploads/2012/08/TAOMA-Brochure.pdf and http://www.fieldsatwork.ch/

**Table 2** Types of measurements and data collected in each INMA study area

|  | Sabadell | Gipuzkoa | Granada | Menorca | Valencia |
|---|---|---|---|---|---|
| Identification of sources |  |  |  |  |  |
|   ELF |  | ✓ | ✓ |  |  |
|   RF | ✓ | ✓ | ✓ | ✓ | ✓ |
| Questionnaires and time-activity diaries | ✓ | ✓ | ✓ | ✓ | ✓ |
| RF geospatial propagation model implemented and validated | ✓ | ✓ | ✓ | ✓ | ✓ |
| Short- and fixed longer-term EMF-NIR measurements |  |  |  |  |  |
|   Indoor |  |  |  |  |  |
|     Homes | ✓ | ✓ | ✓ | ✓ | ✓ |
|     Schools |  | ✓ |  |  |  |
|   Outdoor |  |  |  |  |  |
|     Near homes |  |  | ✓ |  |  |
|     Public playgrounds/parks |  | ✓ |  |  |  |
|     School playgrounds |  | ✓ |  |  |  |
| Personal RF exposure measurements | ✓ | ✓ | ✓ | ✓ | ✓ |

*ELF* extremely low frequency, *RF* radiofrequency of the study population (outdoor –playgrounds/parks– and indoor –homes and schools–), were requested to the pertinent companies (outdoor sources) and to the household members and school teachers (indoor sources). An exhaustive search of all sources that contribute to exposure to EMF-NIR was conducted.

Data requested for environmental outdoor ELF sources, consisted of the presence of high voltage power lines (≥132 kV) and electrical transformation substations (132 kV to 13.2 kV) and stations (13.2 kV to 250–300 V), located within a radius of 200 m from houses, schools and playgrounds. Moreover the railway network map is available for some of the study regions. In the case of outdoor environmental RF sources, mobile phone base stations and radio and television transmitters were identified. All of the appropriate parameters necessary for characterising aforementioned RF sources were requested: location including coordinates (x, y, z), initial and final date of operation, height of the mast (measured from the surface on which it has been installed), type of transmitter, power, communication service, operating frequency, direction (azimuth), vertical orientation (electrical tilt) and number of carrier frequencies. Information regarding indoor environments was collected by questionnaire. Information concerning characteristics (location, height, facade and window frame materials, glazing, etc.), and size of the rooms in which measurements were made and the number and location of sources generating ELF fields (home appliances, music systems, televisions, computers, types of lighting and anti-theft systems) and RF fields (Wireless Access Points for WiFi technology and cordless phone base stations) was gathered. In addition, information on patterns of exposure (places visited by children on weekdays and at weekends, and habits regarding their use of the different abovementioned sources of EMF-NIR) was collected. If children had their own Android smartphone, the XMobiSense app (developed for Android) was installed on the device for a period of at least 4 weeks, in collaboration with the European Project "Characterization of the use of mobile phones in children, adolescents, and young adults" (MOBI-EXPO) [34]. This application measures the real use of mobile phones (number/duration of calls, SMS messages and data transfer), laterality, the use of hands-free controls, and the type of network connection used (2G/3G/4G/WiFi). This data are used to validate the information collected with the questionnaire.

Lastly, parents' perception of the risks associated with exposure to ELFs and RFs was also assessed by questionnaire on a Likert-type scale ranging from 1 to 10.

### Data analysis
Doing a comprehensive analysis and comparing the information gathered on outdoor emission sources and measurement results, the relationship between the proximity to these sources of radiation and exposure levels is analysed. Relationship between indoor exposure levels and data collected through the questionnaire such as the presence and use of electronic and communication devices and characteristics of the buildings (age, materials, number of storeys, etc.) is also studied.

Information on patterns of exposure (time spent in each environment and habits of using EMF-NIR emission appliances) is used along with data obtained with other indirect (geospatial propagation model) or direct methods (environmental and personal measurements) for exposure characterisation of the study participants.

1.2. Implementation and validation of the geospatial propagation model for RFs, NISMap, for the INMA-Spain study area

### Methodology

The collected data on telecommunication transmitters (location, orientation, power, height, etc.), buildings (height, materials, type of windows, etc.) and the 3D environment geometry in each of the study regions are used to construct geospatial propagation models (NIS-Map) to estimate exposure to environmental outdoor RF radiation [24]. Spot measurements of RF fields carried out with ExpoM devices in homes and schools (explained in section 2.1) are used to validate the propagation model for each of the study areas.

### Data analysis

Based on the levels of exposure from the NISMap propagation model, taking into account the time spent in each of the environment (homes, schools, and playgrounds/parks) and having georeferenced their locations, the exposure of 1900 INMA participants to RFs arising from mobile phone, TV and radio transmitters is characterised.

2. Direct methods for assessing EMF-NIR field exposure
    2.1. Spot and fixed longer-term environmental measurements of EMF-NIR

### Methodology

Spot and fixed longer-term measurements in ELF and RF ranges (5 Hz to 6 GHz) inside homes (children's bedroom and living/dining room) and schools (classrooms) and outdoors (school playgrounds, public playgrounds/parks, and areas around the homes) were carried out (Table 3). The measurement methodology varies as a function of the type of field, frequency, and environment (indoors or outdoors), as well as the type of measurement equipment used.

Below, we describe some key aspects of some of these measurements that are not fully explained in Table 3.

Longer-term (17 to 24 h) measurements of ELF fields in homes (indicated in the column "Type of EMF-NIR" in the Table 3 by a superscript 1) consisted of initially placing the probe in the middle of the living/dining room, then moving it to the middle of the bedroom when the child went to bed and moving it back to the living/dining room in the morning, recording the times at which the device was moved [33].

Further, to identify the contribution of different RF sources (frequencies) to the total radiation exposure, spot (2 min) measurements of RF fields were carried in houses and classrooms (indicated in the column "Type of EMF-NIR" of the table by a superscript 2), following the procedure described by Röösli et al. [28] and the European Committee for Electrotechnical Standardization [35]. Inside the houses, these measurements were made with doors and windows closed and when people were not present, to minimise potential interference, whereas at schools, they were carried out while the children were in the classrooms to avoid the inconvenience of taking them out of the room and to obtain exposure levels which correspond to school hours when the children are attending the class. The tracking procedure used for measuring RFs in public playgrounds/parks (indicated in the column "Type of EMF-NIR" of the table by a superscript 3) involved moving across the whole area along a zigzag path at a constant speed to obtain a mean level of radiation and identify the points of maximum exposure. At the points where the highest levels were detected, we took an additional 20-min measurement to differentiate spot peaks from usual levels.

### Data analysis

All direct measurements (spot and fixed longer-term) obtained in the houses, schools and playgrounds/parks are used to characterise the exposure to EMF-NIR in the different locations where the participants spend most of their time and to identify the locations that contribute most to their exposure. Measures of central tendency, such as arithmetic and geometric means, as well as the quadratic mean (RMS), standard deviation, median, and range of exposure to EMF-NIR (5 Hz-6 GHz) in each environment are calculated. In indoors environments, where spot measurements are done both in the centre and corners the mean for each room was calculated taking into account all the measurements. Moreover, for each of the environments analysed, spectral analysis of the exposure levels are carried out, from the data obtained with the devices measuring narrowband radiation (NARDA EHP-50D and ExpoM-RF 3). In this way, the value of each of the spectral components, that is, the different frequency bands that contribute to the total level is ascertained, and the main sources of ELF and RF emissions are identified. In addition, longer-term measurements are used to assess spatial and temporal variability in levels of exposure.

Further, with the exposure levels calculated for each environment, together with the data on time-activity patterns from the questionnaires, levels of individual exposure are estimated, considering the time children spend in each location. In this way, individual level of exposure to EMF-NIR in the 5 Hz to 6 GHz frequency range from environmental (internal and external) sources is assigned to a subsample of 200 children (see Table 3). This methodology does not consider exposure due to the

**Table 3** Spot and fixed longer-term measurements of electromagnetic fields of non-ionising radiation (EMF-NIR)

| | N | Type of EMF-NIR | Field | Frequency range | Duration (measurement interval) | Probe height | Measurement site | Measurement equipment |
|---|---|---|---|---|---|---|---|---|
| INDOOR | | | | | | | | |
| Homes | 123 | ELF[1] | EF/MF | 15 Hz – 100 KHz | 17 h (240) | 79 cm | Middle | Taoma |
| Bedroom & living/dining room | | | | | | | | |
| Homes & schools | 80[a] & 26[a] | ELF | EF/MF | 3 spans (100 Hz[b], 1 KHz, 100 KHz)[c] | Spot measurements | 1.1 m | Middle & each corner[d] | Narda |
| Bedroom, living/dining | | ELF[1] | | 6 - 500 Hz | 24 h (30 s) | 1.1 m | | Narda |
| room & classroom | | RF | MF | 100 KHz – 6 GHz | 2 min (1 s) | 1.1 m | Middle | Narda |
| | 300 | RF[2] | EF/MF | 16 bands (87.5 MHz-5.8 GHz)[e] | 2 min (4 s) | 1.7. 1.5 & | Middle & each corner[d] | ExpoM |
| | | | EF | | | 1.1 m[f] | | |
| OUTDOOR | | | | | | | | |
| Near homes | 123 | RF | EF | 100 KHz – 6 GHz | 6 min (1 s) | 1.45 m | 2 m from the home | Taoma |
| Public playgrounds/parks | 151 | ELF | MF | 6-500 Hz | 20 min (30 s) | 1.1 m | Middle | Narda |
| | | ELF | EF/MF | 3 spans (100 Hz, 1 KHz, 100 KHz)[c] | Spot measurements | 1.1 m | Middle | Narda |
| | | RF[3] | EF/MF | 100 KHz – 6 GHz | Tracking | 1.1 m | Whole area | Narda |
| | | RF | EF/MF | 100 KHz – 6 GHz | 10 min (1 s) | 1.1 m | Middle | Narda |
| | | RF | EF | 16 bands (87.5 MHz - 5.8 GHz)[e] | 6 min (4 s) | 1.1 m | Middle | ExpoM |
| | 25 | ELF | MF | 6 - 500 Hz | 20 min (30 s) | 1.1 m | Middle | Narda |
| School playgrounds | | ELF | EF/MF | 3 spans (100 Hz, 1 KHz, 100 KHz)[c] | Spot measurements | 1.1 m | Middle | Narda |
| | | RF | EF/MF | 100 KHz – 6 GHz | 20 min (1 s) | 1.1 m | Middle | Narda |
| | | RF | EF | 16 bands (87.5 MHz - 5.8 GHz)[e] | 6 min (4 s) | 1.1 m | Middle | ExpoM |

*RF* radiofrequency, *ELF* extremely low frequency, *EF* electric field, *MF* magnetic field; [a]In Gipuzkoa 80 homes and 26 schools have been measured for the whole EMF-NIR range; [b]Only in the middle; [c]The start frequency for each span corresponds to 1.2 % of the span. The frequency bands are: 5–100 Hz; 12 Hz-1 kHz; 1.2-100 kHz; [d]At 1.4 m from the wall; [e]Frequency bands: 87.5 - 108 MHz; 470–790 MHz; 791 – 821 MHz; 832 – 862 MHz; 880 – 915 MHz; 925 –960 MHz; 1710 – 1785 MHz; 1805 – 1880 MHz; 1880 – 1900 MHz; 1920 – 1980 MHz; 2110 – 2170 MHz; 2400–2485 MHz; 2500 – 2570 MHz; 2620 – 2690 MHz; 3400 – 3600 MHz; 5150 – 5875 MHz; 1,2 and 3; methodology described in more detail in the text; [f]At 1.7, 1.1 and 1.5 m in the middle, and only at 1.5 m in the corners personal use of information and communication technology devices or other electrical appliances.

### 2.2. Personal measurements of children's and adolescents' individual RF exposure

#### Methodology

Individual measurement to EMFs across 16 frequency bands (between 87.5 MHz and 5.87 GHz) were obtained in 300 children from all study regions, using ExpoM-RF 3 personal RF exposimeters with a measurement interval of 4 s. The procedure consisted of children wearing the exposimeter up to three consecutive days (72 h), the device being placed in a padded belt bag, with no metal items. They were advised to wear the bag around the waist when possible during the day. At night, children placed the exposimeter on a flat non-metallic surface, as close as possible to their bed. The exposimeters used had a GPS which provides data on the location of the participant at all times. Participants or, in the case of children their parents, also completed an activity diary recording their schedule during measurements days and a questionnaire concerning the measurement period asking them about: their activities and places they had been; the place they had kept the belt bag with the exposimeter most of the time (rucksack, waist, etc.), and whether, and if so for how long, they had used RF sources (mobile phone, and if so, where they kept it, cordless phone, computers, tablets and videogames with Internet via WiFi, 3G/4G or cable). In a subsample of 30 children we performed a repeatability study one year later following the same protocol.

### Data analysis

Data on personal measurements will be combined with the information recorded in the activity diary in order to provide additional relevant information related to RF exposure levels of 16 different frequency bands in different environments and its spatial and temporal variability, as well as the sources of emission and activities of the children that contribute most to their RF exposure in outdoor and indoor environments. Moreover, the repeatability study will enable to investigate the reproducibility and temporal variation of RF exposure in children one year later.

## Discussion

The methodology proposed in this paper will enable the characterisation of exposure to EMF-NIR in children and adolescents, including i) ELFs and RFs by means of direct methods in around 200 participants (spot and fixed-long term measurements), ii) total RF exposure, including environmental and personal exposure, in 300 participants by means of direct methods (personal measurements) and iii) outdoor environmental RF exposure in 1900 participants by means of indirect methods (NIS-Map modelling). This represents a sound base for future research into potential effects of EMF-NIR exposure on health, such as neuropsychological development and contributes to the current knowledge on the characterisation of EMF-NIR exposure [36].

Multiple factors are involved in the interactions between children environment and health. The use of EMF-NIR sources is gradually increasing among children and adolescents, and hence, it is important to consider this type of exposure in cohort studies investigating the effects of exposure to environmental hazards. The INMA Project has previously collected data on prenatal and postnatal exposure to several chemical pollutants and has assessed the association between this exposure and the growth and development of children at different childhood stages. The evaluation of exposure to different sources of EMF-NIR extends the characterisation of environmental exposures in this cohort.

Many epidemiological studies have characterised EMF-NIR exposure partly, and several have assessed the effects of some sources of EMF-NIR on health. To investigate the relationship between EMF-NIR exposure and potential effects, it is essential to carry out a proper characterisation of this exposure. The most common criticisms of studies describing effects associated with

EMF-NIR are the methodology used for characterise exposure and failing to take into account most of the sources of emission [16, 17]. In particular, exposure to ELFs has generally been estimated by indirect methods such as those described by Wertheimer and Leeper [4]. However, studies are starting to emerge in which ELF exposure is estimated by spot and longer-term direct measurements, and taking into account all the potential sources of emission [33], while others use personal measurements [37]. With regards to RFs, the majority of studies published on the effects of RF exposure on health consider the use of mobile and cordless phones (900 MHz to 2600 MHz), with data mostly from questionnaires [6–8, 38], not taking into account environmental sources of RF EMFs (mobile phone base stations, radio and TV transmitters, WiFi and cordless phone base stations), to which children are exposed on a daily basis, at school and/or at home. These studies analyse the effects of the exposure to RF radiation from devices operated close to the body, generally to the head, from time to time and for short periods. However, radiation from RF environmental sources tends to be more homogeneous and weaker than that from the aforementioned sources, but exposure involves the entire body and continues for longer periods of time.

In our study, indirect and direct methods have been used for the characterisation of the exposure to EMF-NIR. Regarding indirect methods, information was collected on the proximity of different environments (houses, schools and parks) to RF transmitters and electricity transmission and distribution lines and on the use of ELF and RF emission sources by questionnaires. However, to avoid the limitations of the use of this type of indirect method, the gathered information is not used on its own for the exposure assessment but to complete and improve the characterisation done using other methods. Furthermore, using the XMobiSense app, we have overcome the problems related to assessing mobile phone use with questionnaires, which tend to underestimate the number of calls made/received and overestimate the length of the calls [34, 39]. Although we will be able to obtain a good estimate of the real use of mobile phones, this will only be possible in the case of the oldest children with their own smartphone. Moreover, we will not count with the information on the output power of the mobile phone, which is essential for the exposure characterisation. In order to solve it, the participants provide information regarding the specific brand and model of the mobile phone they use most.

Within indirect methods, geospatial propagation models may be a good alternative to direct methods estimate environmental exposure of the population to RFs arising from mobile and radio/TV base stations, since they significantly reduce the need for materials and overall costs. However, they require high quality data to be available on the technical specifications of the emitters, as well as the geospatial characteristics of buildings in the area. Further, this methodology does not allow to estimate the level of exposure to non-environmental sources such as mobile phones and other wireless communications systems, including WiFi or cordless phone base stations. On the other hand, provided that technical information from the emitters and land registry is available, the NISMap model is a useful tool since it makes it possible to estimate retrospective exposures.

To our knowledge, models have not yet been developed and validated to predict ELF exposure. For this reason, characterisation of exposure to radiation in this frequency range relies on direct methods, as well as the collection of information on environmental sources of exposure such as high- and medium-voltage power lines and others, including home appliances.

For reliable characterisation of exposure using direct methods, the selection of measurement devices is of great importance, especially with respect to the limit of quantification. We have used Narda and Taoma devices to measure broadband RFs, which have measurement ranges (0.375 to 650 V m$^{-1}$ and 0.2 to 340 V m$^{-1}$ respectively) that enable us to check whether exposure levels are within legal limits [40], but many of the fields measured with these systems were below the limit of quantification (0.375 and 0.2 V m$^{-1}$) [32]. This limitation has been partially overcome by use of the personal narrowband exposimeter (ExpoM-RF 3) which has a lower limit of quantification (0.003 to 0.02 V m$^{-1}$, depending on the frequency band). This allows us to determine, on the one hand, children's exposure levels with a greater accuracy, and on the other, the types of sources that contribute most to these levels. Another important point to consider is the height at which measurements are taken to enable comparisons between the results of different studies. In relation to this, the Institute of Electrical and Electronics Engineers recommends measuring RF fields at a height of up to 2 m [41] and ELF fields at 1 m [42] above the floor. We measured RFs at 1.1, 1.45, 1.5 and 1.7 m and ELFs at 0.79 and 1.1 m above the floor, considering the different ages and heights of the participants involved (7 to 18 years old).

Spot and fixed longer-term measurements make possible to carry out the estimation of exposure to the full range of EMF-NIR, including ELF and RF (5 Hz to 6 GHz). However, the frequency ranges at which emitters are operated and spatial and temporal variability in emissions govern the levels of indoor and outdoor exposure, and in turn, individual exposure fluctuates depending on these factors. Hence, according to some authors, the estimates obtained from spot and fixed longer-term measurements may not be fully representative of personal exposure levels [16]. Another disadvantage of this methodology is the great effort required in terms of time and resources, and for this reason, we are only making estimates of both ELF and RF exposure from direct measurements in a subset of 200 children from the cohort. Resource limitations mean that we will only be able to carry out a comprehensive characterisation of EMF-NIR exposure for children in whose homes measurements of the whole frequency range have been made.

Estimating individual levels of RF exposure using a personal exposimeter (ExpoM) should provide us with more realistic data. However, this methodology has some limitations in that it may alter the real exposure as reported by Frei et al. [43], due to shielding effects or potential variations in the normal behaviour of children when using the device.

Besides our study, other epidemiological studies, such as the HERMES [30] and the ABCD [27, 44] cohort studies in Switzerland and the Netherlands respectively, the cross-sectional MoRPhEUs study in Australia [11] and the multicentre case control CEFALO study in Scandinavian countries and Switzerland [8] have characterised RF exposure. With the exception of ABCD cohort, the rest were created with the aim of assessing the exposure and effects of RF fields. As well as the INMA cohort study, other European cohorts that are involved in the GERoNiMO Project (ABCD, HERMES and the Danish DNBC child cohort [45]) will have information on RF exposure along with data on other covariates. However, to date, we have not found any cohort studies that explore exposure to the whole range of EMF-NIR frequencies and that also provide information on exposure to a large number of environmental pollutants and covariates such as that analysed in the INMA cohort study.

To our knowledge, there is just one previous study comprehensively characterising RF exposure that has also assessed its effects on neuropsychological development [46]. Research in this field is limited to challenge studies using volunteers subjected to acute exposure to certain RFs [47–49]. Regarding exposure to the magnetic component of ELF fields, most studies have focused on a potential carcinogenic effect [50], although some have also explored adverse effects on cognitive functions in children [51, 52].

In conclusion, our work will contribute to understanding the main sources of EMF-NIR exposure in children and adolescents at different ages (from 7 to 18 years old) and their contribution to exposure in daily life, since these may differ from patterns in adults [29, 53, 54] or in adolescents from other countries [30]. Such information is essential to assess the relevance of each source of exposure to child development and together with other project data, on socioeconomic and family characteristics, and children's health, will allow us to assess more comprehensively, in later stages of follow-up, the complex interaction between the children's health and development and various environmental factors that surround them.

Finally, given advances and changes in technology and constantly changing patterns in its use, exposure to EMF-NIR should be studied continuously at different stages during childhood, to improve our understanding of their real exposure and its potential effects on their health, as well as analyse the relevance of cumulative exposure over time.

### Abbreviations
ELF: Extremely Low Frequency; ELF-EF: Extremely Low Frequency-Electric Field; ELF-MF: Extremely Low Frequency-Magnetic Field; EMF: Electromagnetic Fields; EMF-NIR: Electromagnetic Fields of Non-Ionizing Radiation; GERoNiMO: Generalised EMF research using novel methods. An integrated approach: from research to risk assessment and support to risk management; IF: Intermediate Frequency; INMA: Environment and childhood (from *infancia y medio ambiente*) cohort; Mobi-Expo: Characterization of the use of mobile phones in children, adolescents, and young adults; Rembrandt: Radiofrequency ElectroMagnetic fields exposure and Brain Development; RF: Radiofrequency.

### Competing interests
The authors declare that they have no competing interests.

### Authors' contribution
MGa, AJ, LS and MGu wrote the manuscript. All authors contributed to the design of the study. DG, MF and JG provided technical advice. MGu coordinated the work done in the INMA study under the European projects GERoNiMO and REMBRANDT. All authors read and approved the final manuscript.

### Acknowledgments
We would like to thank all INMA families for their active collaboration, this being essential for achieving the objectives of the INMA Project.
MGa would like to thank the Department of Education, Language Policy and Culture of the Government of the Basque Country for a predoctoral research training grant.
The characterisation of the exposure to non-ionising electromagnetic radiation in children of the INMA Gipuzkoa cohort has been funded by the Spanish Ministry of Economy and Competitiveness (FIS) and by the councils of the study areas.
This study was funded by grants from the Spanish Instituto de Salud Carlos III Health Institute (PI13/02187 incl. FEDER funds, CP13/00054 incl. FEDER funds, MS13/00054), the councils of the study region of Gipuzkoa and the EU Commission (603794).

### Ethical declaration
Prior to children's inclusion in the study, their legal guardians provided written informed consent. The research has been performed in accordance with the Spanish Law 14/2007 on Biomedical Research and the ethical principles of the Declaration of Helsinki. The work carried out under the REMBRANDT Project in all the cohorts has been approved by the ethical committee of the CEIC-Salut de Parc Salut Mar. The work carried out in Gipuzkoa and Granada has been approved by the ethical committees of the Basque Country (CEIC-E) and of San Cecilio University Hospital of Granada, respectively.

### Author details
[1]BIODONOSTIA Health Research Institute, Paseo Dr. Beguiristain, San Sebastian 20014, Spain. [2]University of the Basque Country (UPV/EHU), Faculty of Pharmacy, 7 Unibertsitateko Ibilbidea, Vitoria-Gasteiz 01006, Spain. [3]ISGlobal, Centre for Research in Environmental Epidemiology (CREAL), Barcelona Biomedical Research Park, C/Doctor Aiguader 88, 08003 Barcelona, Spain. [4]Pompeu Fabra University, C/Doctor Aiguader 88, 08003 Barcelona, Spain. [5]Spanish Consortium for Research on Epidemiology and Public Health (CIBERESP), Instituto de Salud Carlos III, C/Monforte de Lemos 3-5 28029 Madrid, Spain. [6]Department of Child and Adolescent Psychiatry/Psychology, Erasmus University Medical Centre–Sophia Children's Hospital, PO Box 2060, 3000 CB Rotterdam, The Netherlands. [7]Public Health Division of Gipuzkoa, Basque Government, 4 Av. de Navarra, San Sebastian 20013, Spain. [8]University of Granada, San Cecilio University Hospital, Instituto de Investigación Biosanitaria (ibs.GRANADA), Granada 18071, Spain. [9]Communications Engineering Department, University of the Basque Country (UPV/EHU), Faculty of Engineering, Alameda Urquijo, Bilbao 48013, Spain. [10]Swiss Tropical and Public Health Institute, Socinstrasse 57, Basel 4002, Switzerland. [11]University of Basel, Basel, Switzerland. [12]Epidemiology and Environmental Health Joint Research Unit, FISABIO–Universitat Jaume I–Universitat de València, Avenida de Catalunya 21, Valencia 46020, Spain. [13]IB-Salut Menorca Health Area, Balearic Islands, Spain. [14]University of the Basque Country (UPV/EHU), Faculty of Medicine, San Sebastian, Spain. [15]Materials Physics Department, University of the Basque Country (UPV/EHU), Faculty of Chemistry, Paseo Manuel de Lardizabal 3, San Sebastian 20018, Spain.